\documentclass{article}
\usepackage{arxiv}
\usepackage[utf8]{inputenc} 
\usepackage[T1]{fontenc}    
\usepackage{hyperref}       
\usepackage{url}            
\usepackage{booktabs}      
\usepackage{amsfonts}       
\usepackage{nicefrac}       
\usepackage{microtype}     
\usepackage[norelsize]{algorithm2e}
\usepackage{multirow}
\usepackage{tabulary}
\usepackage{graphicx}
\usepackage{subcaption}

\title{HAUAR: Home Automation Using Action Recognition}

\author{
  Shashank Kotyan, Nishant Kumar, Pankaj Kumar Sahu, and Venkanna U.
  \thanks{Department of Computer Science, Dr. Shyama Prasad Mukherjee International Institute of Information Technology, Naya Raipur (IIIT-NR), Chhattisgarh, India}
  \\
  \texttt{\{shashank15100, nishant15100, pankajs15100, venkannau\}@iiitnr.edu.in} \\
}

\begin{document}
\maketitle

\begin{abstract}
Today, many of the home automation systems deployed are mostly controlled by humans. This control by humans restricts the automation of home appliances to an extent. Also, most of the deployed home automation systems use the Internet of Things technology to control the appliances. In this paper, we propose a system developed using action recognition to fully automate the home appliances. We recognize the three actions of a person (sitting, standing and lying) along with the recognition of an empty room. The accuracy of the system was 90\% in the real-life test experiments. With this system, we remove the human intervention in home automation systems for controlling the home appliances and at the same time we ensure the data privacy and reduce the energy consumption by efficiently and optimally using home appliances.
\end{abstract}


\section{Introduction}
Computer vision (CV) is the branch of computer science which deals in determining the features and attributes of the image(s). These features and attributes of an image provide meaningful inferences concerning specific applications. The applications of CV have grown to almost all the domains of computer science which involves analysis and interpretation of data and image, since the past 30 years \cite{Walsh}. Among, all applications, home automation (HA) is emerging application with growing importance. \\
Human civilization is in the constant growth and exploration of automating things in every sphere of life whether it is automating home appliances, automating the computations, automating vehicles, automating learning, and all such automation \cite{Hossai}. A smart home can be defined as entwinement of various elements like sensors, connections, and applications. These elements together build a dynamic heterogeneous architecture. HA means to monitor and automate the utilization of home appliances while maintaining the security. HA makes the human life simpler and more relaxing along with efficient usage of the home appliance. HA relieves us from the worry of turning on the fan after a long journey out of tiredness or worry about the temperature of the room when there are many people in the room. Most importantly, it relieves us from the worry of turning off the home appliances when we are not around. HA is an important area of research in the era of automating things and has found immense interest among researchers. This interest has bloomed since Internet of Things (IoT) has become a potential field of research in the recent years, where all the devices communicate with other. The aim of a smartphone in HA is to efficiently manage home appliances with advanced or sophisticated services provided to the users \cite{Geneiatakis}. It has also been estimated that around 90 million people all over the world will be living in smarter homes in the coming years \cite{Jacobsson}. Research in HA has steadily progressed and over the period, home networks have become a necessity in a smart home. HA’s research is not limited to academic research but has expanded to the attention of industries who are continually trying to use the home appliances optimally for better efficient energy usage. The common motivation behind these research is to develop a fully automated home automation system which is a potential challenge. There is no state of the art practice to design a system for home automation. The primary concern includes energy consumption by home appliances. IoT which can be used to reduce the energy consumption by home appliances remotely without the physical presence of the user is also a challenging task \cite{Deore}. With human intervention, home appliances are not utilized efficiently and there is a lot of energy wasted. The problem of preventing home appliances to consume energy in the absence of human is one of the fundamental problems which is dealt with almost all HA systems. As the number of home appliances per person is increasing so is the energy consumption. Keeping a check on energy consumption has also found some immense industrial attention.\\
The motivation to develop an intelligent HA system has grown from two basic needs of human. First, the hunger to look for more comfort using the existing technologies. Second, in the environment conscious era to reduce the energy consumption. The current HA systems are mostly based on the smartphonebased control technology, which is a type of human intervened system. A human intervened system has its own set of problems, in efficiently monitoring and controlling the home appliances. It also puts a limit on the automation of home appliances which refrains us from optimally utilizing the appliances. Also, the current HA system incorporates the user who is aware of the smartphone technology and is proficient in using it, but this limits the coverage of audience. Our prototype, therefore, has grown to meet these basic needs of human. It provides more comfort to the user with making home appliances hassle free and automated. Along with it, it also optimally uses the home appliances to reduce the wasteful energy consumption which arises in case of manual controlling of home appliances. The problem stuck to us when we observed the utilization of lights, fans and air conditioners in the absence of a human. This was happening because of mainly two reasons, one was human ignorance where people do not bore the sense of authority over the appliances and let it run or, wait for some other person to turn off the appliances. Another reason was that of miscommunication which happened between the people, where people thought that the other person would turn off the appliances. \\
The proposed prototype is an intelligent prototype capable of taking decisions without any human intervention and take care of concerns like data privacy and energy consumption with diligence. This prototype will detect a person based on its motion and state of a person from standing, sitting and lying based on an image captured and accordingly will take necessary actions to turn on the home appliances and control their intensity. The prototype also incorporates temperature and humidity sensor to increase the comfort of the user. 

\section{Related Works}
Present HA systems and projects involve the intervention of human to automate the home appliances. This intervention can be provided by the use of mobile applications, EEG signals from the brains or web applications. Some of the existing solutions are described below. Bentabet et al. used EEG signals from the brain of the user by staring at a screen to monitor and somewhat automate the home appliances. This method was particularly developed for the paralyzed people for whom simple tasks such as switching on and off the light is a challenging task. Brain-Computer Interface (BCI) is an emerging technology which provides direct communication between the computer and the human brain, without resorting to conventional pathways. This allows performing various kinds of tasks and integrates disabled people into society by using their cerebral or brain activities. The BCI system takes EEG signals as input and process it i.e., classify it by extracting features. Then it translates these predictable signals into commands to control the system and perform the necessary actions. Feedback is necessary for this approach to verify that the user's intent was correctly executed. This approach required user focus on a visual or tactile stimulus. However, EEG signal strength varies from person to person and person with high focus skill-set transmit better EEG signal to use the home appliance accurately \cite{Bentabet}. Kumar et al. propose the use of Bluetooth as a suitable means for controlling home appliances. They have used both Raspberry Pi (RPi) and Arduino for their prototype. It uses HA system as a substitute for an electrical switch. Arduino processed the control signal for appliances as well as the text message of the communication system \cite{Kumar}. Deore et al. implemented a home controlling system using Arduino and a smartphone application to control appliances such as lights, motor, garden irrigation, and a timer for home appliances and gate \cite{Deore}. Patchava et al. created a practical home automation system using, camera, motion sensors and it streamed images captures by the camera using M-JPG Streamer to a web application which allowed the users to control the appliances from anywhere in the world using an internet connection \cite{Patchava}. Peng et al. pointed the limitations of network enabled devices which are common in IoT. One technical problem was a limitation to provide intelligent and flexible services. The other was cooperation between various heterogeneous IoT devices and their role allocation \cite{Peng}. Another drawback of Wireless Sensor Networks (WSNs) based systems is scalability and maintainability \cite{Baroffio}. These modules though monitor the home appliances remotely, but they do not automate the home appliances as the input has to come from the user to control the home appliances. One is a hardware limitation of Infrared-based systems which have a small range of about 5 meters \cite{Sefat}. Tsai et al. proposed a method for HA using object detection and recognition. It mainly focused on height recognition as height is highly correlated to the detected object. The object is recognized as either an adult or child with the help of height of the object \cite{Tsai}. Trade-off which comes from deploying visual sensor networks is of data generation and transmission which impacts bandwidth usage and energy consumption. Transmission of JPEG encoded image over a communication channel is an expensive operation which also introduces a significant amount of latency but if we decrease the quality of the image or decrease the histogram resolution, then we also affect the classification accuracy \cite{Baroffio}. Shinde et al. proposed home automation using gesture recognition. Few limitations of gesture recognition are complicated backgrounds, obstructions and static and dynamic hand gestures. This gives rise to this sophisticated analysis of gestures. Recognizing scaled, translated and rotated hand gestures is computationally inefficient and is a rigorous process \cite{Shinde}. Sefat et al. proposed a system for home security system using CV due to defects in sensor-based approaches. One technique of automatic counting of people using light beam suffered the inconsistent result when more than one people pass the baseline in parallel due to which it cannot be accurately told how many people crossed the beam. The limitation of adopting computer vision over traditional sensor networks which came were lighting and shadows. Lighting affected the camera images drastically and ineffective choice of lightening conditions caused trouble in implementation. Also, inappropriate lightening led to false results in many cases and decreased the accuracy of the system. The shadow which was captured also contributed to providing inefficient results \cite{Sefat}. Present Home automation with the use of CV also have limitations of energy consumption and false results due to occlusion, overlapping, and shadow effect. Deore et al. noticed that the well-established HA systems are based on wired communication, and most home automation systems require human interaction through applications which restrict the flexibility and scalability of the system. Other relevant problems to home automation systems were of costs, manageability, security and energy consumption \cite{Bentabet}. These human interventions in monitoring the home appliances limit the automation according to the usage of humans and cannot be called as intelligent. An abstract idea of HA is automating the home appliances according to the needs of a person which are determined without the intervention of the person. Solutions to various problems in different domains have been resolved by introducing CV which yields a better performance and accuracy. Also, they are easy to deploy as they can use the existing visual sensor networks. CV is already implemented in various domains like parking systems replacing the traditional sensor networks by overcoming the limitations of sensor networks \cite{Shinde}. 

\section{HAUAR: Home Automation Using Action Recognition}
\begin{figure}
  \centering
  \includegraphics[width=.99\linewidth]{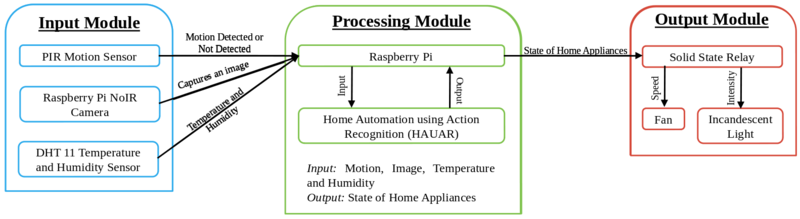}
  \caption{System Design consisting of input, processing and output module}
  \label{fig:system}
\end{figure}

\begin{figure}
  \begin{subfigure}{.5\textwidth}
    \centering
    \includegraphics[width=.8\linewidth]{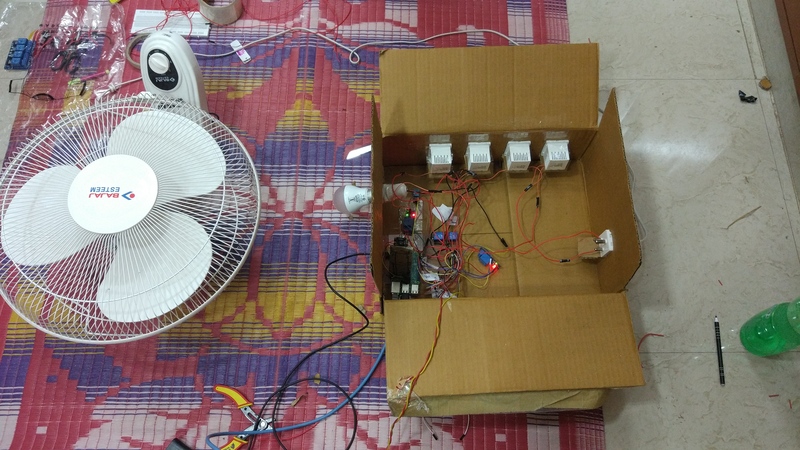}  
    \caption{}
    \label{fig:bird_a}
  \end{subfigure}
  \begin{subfigure}{.5\textwidth}
    \centering
    \includegraphics[width=.8\linewidth]{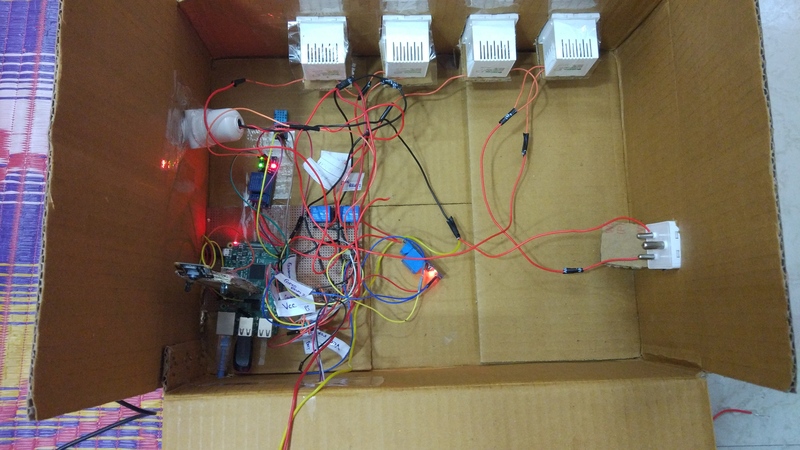}  
    \caption{}
    \label{fig:bird_b}
  \end{subfigure}
  \caption{Bird view of our proposed HAUAR prototype }
  \label{fig:bird}
\end{figure}

Our proposed prototype’s design can be split into three distinct individual modules (input, processing, and output) as shown in Figure \ref{fig:system}. The modules involved in the prototype are explained further in the section. Figure \ref{fig:bird_a} shows the bird view of the prototype developed implementing the proposed algorithm while Figure \ref{fig:bird_b} shows the bird view of the processing module of the prototype. \\

\subsection{Input Module}
The input module takes care of perceiving the environment which consists of three components,
\emph{Pyro-electric Infrared sensor}: A Pyro-electric Infrared (PIR) sensor is an electronic device, which is mainly used for motion detection. It is a three pin device (ground, signal, and power). It can detect the motion of human being within the range of 10 meters (approximately). This device is used in many IoT projects where the motion detection is required.
\emph{Raspberry Pi NoIR camera}: Raspberry Pi NoIR camera module is used for taking the picture and recording the high definition video. It is available in different specifications (5MP, 8MP, etc.). Pi camera module provides a particular port in the raspberry pi and works with a simple command after enabling. In the NoIR camera, there is no infrared filter on the lens due to which it is capable of taking a picture in low light or infrared illumination. 
\emph{DHT 11 temperature and humidity sensor}: It is a three pin device (ground, signal, and power). It is a low-cost and can be interfaced with any microcontroller like RPi, Arduino, etc.

\subsection{Processing Module}
The processing module takes care of the processing of the perceived data from the input module to convert it into information. This module consists of RPi, which runs various algorithms for processing. RPi is a mini computer with 1 GB RAM with the size of an ATM card. It has four USB port for connecting mouse, keyboard, and transferring the data, micro SD Card slot for booting the system, an Ethernet port for connecting the internet, an audio jack for sound, a display port, a camera port for connecting the Pi camera, a HDMI port and a 12V DC power supply port. It also provides Wi-Fi and Bluetooth connectivity. It is mainly used for educational purposes and academic projects of small scale. In the proposed algorithm, we first take the input from PIR sensor persistently to detect a motion. Whenever a motion is detected or time stamp takes the value of 30 minutes, room’s temperature and humidity are read through DHT 11 sensor and an image of the room is captured by the Raspberry Pi NoIR Camera, the image is then preprocessed. The pre-processed image is then set to input for HOG classifier and HAAR cascade classifier to detect the presence of a human. This is done to prevent any false which operates at 3-5V power. The output of the sensor can be positive results from PIR and also to partition the images into multiple segments when there is more than one person. These segmented images along with the original image are passed to the inception to classify the image as either empty room, sitting, standing or lying. An empty room is a classifying label because HOG classifier and HAAR cascade classifier have a high chance of giving false positive results which can be seen in the images in the results section. After classifying, the state of the home appliances is decided according to the state table. Table \ref{table:state} mentions the various states possible for our prototype to act. We predefine our prototype with 13 states based on temperature, humidity, and occupancy of the room.

\begin{table}
  \caption{State Table for Home Appliances}
  \centering
  \begin{tabular}{lllll}
    \toprule
    \multicolumn{3}{c}{State}             & \multirow{2}{*}{Fan}  & \multirow{2}{*}{Incandescent Light} \\
    \cmidrule(r){1-3}
    Occupancy   & Temperature & Humidity  &             &                   \\
    \midrule
    Empty Room  & Any         & Any         & Off           & Off                           \\
    \midrule
    Standing    & Low         & Low         & Low           & Bright                        \\
    Standing    & Low         & High        & High          & Bright                        \\
    Standing    & High        & Low         & High          & Bright                        \\
    Standing    & High        & High        & High          & Dim                           \\
    \midrule
    Sitting     & Low         & Low         & Low           & Bright                        \\
    Sitting     & Low         & High        & High          & Dim                           \\
    Sitting     & High        & Low         & High          & Bright                        \\
    Sitting     & High        & High        & High          & Dim                           \\
    \midrule
    Lying       & Low         & Low         & Low           & Dim                           \\
    Lying       & Low         & High        & Low           & Dim                           \\
    Lying       & High        & Low         & High          & Dim                           \\
    Lying       & High        & High        & High          & Dim                           \\
    \bottomrule
  \end{tabular}
  \label{table:state}
\end{table}

The following is the pseudo code of the algorithm, we propose the algorithm using the pseudo code described in Algorithm \ref{algo:hauar}.

\begin{algorithm}[H]
\caption{HAUAR}
\SetAlgoLined
\KwIn{Motion, Timestamp}
\KwResult{State}
  \eIf{Motion = True or Timestamp = 30min}{
    Read temperature and humidity of room\\
    Capture an image of the room\\
    Pre-process the image\\
    Detect faces, lower body, upper body, full body using haar
cascades on image\\
    Detect people using HOG and SVM on image \\
    Segment region of interest\\
    Run customised inception CNN with all segments containing
region of interests\\
    Predict number of standing people, sitting people and lying
people or empty room \\
    Deduce state of home appliances using state table \\
    Timestamp = 0 \\
    }
    {
      Do Nothing
    }
  \Return{State}
\label{algo:hauar}
\end{algorithm}

\subsection{Output Module}
The output takes care of the reflecting back the action back to the environment in the form of controlling the home appliances. This module consists of home appliances connected through relay switch which are meant to be automated using The output takes care of the reflecting back the action back to the environment in the form of controlling the home appliances. This module consists of home appliances connected through relay switch which are meant to be automated using HAUAR. RPi operates at 5 volts, but home appliances require 220/230 volts for usage therefore to control the home appliances we require a relay. A relay is an electromagnetic switching device which is mainly for controlling devices which operate at higher voltages than RPi. The functioning of the relay is just to break the current and re-establish the connection.

\section{Experiments}
\begin{figure}
  \begin{subfigure}{.33\textwidth}
    \centering
    \includegraphics[width=.8\linewidth]{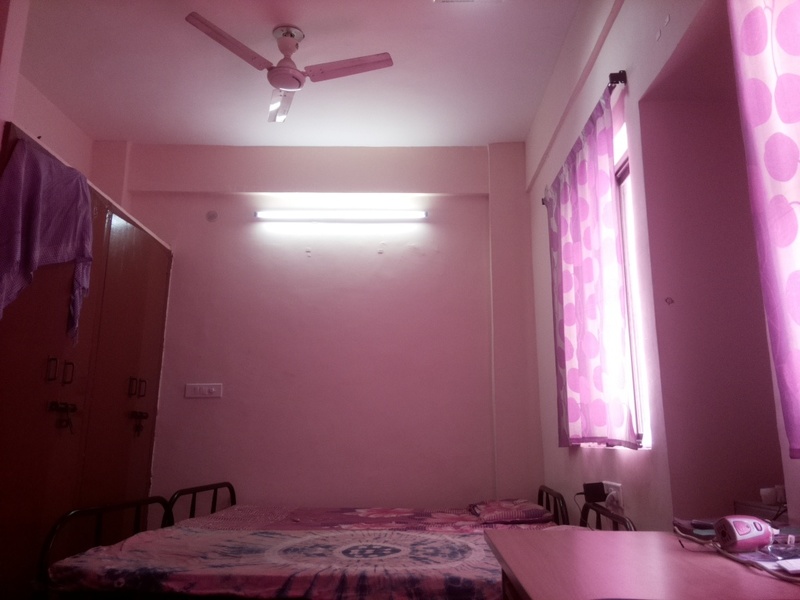} 
    \caption{} 
  \end{subfigure}
  \begin{subfigure}{.33\textwidth}
    \centering
    \includegraphics[width=.8\linewidth]{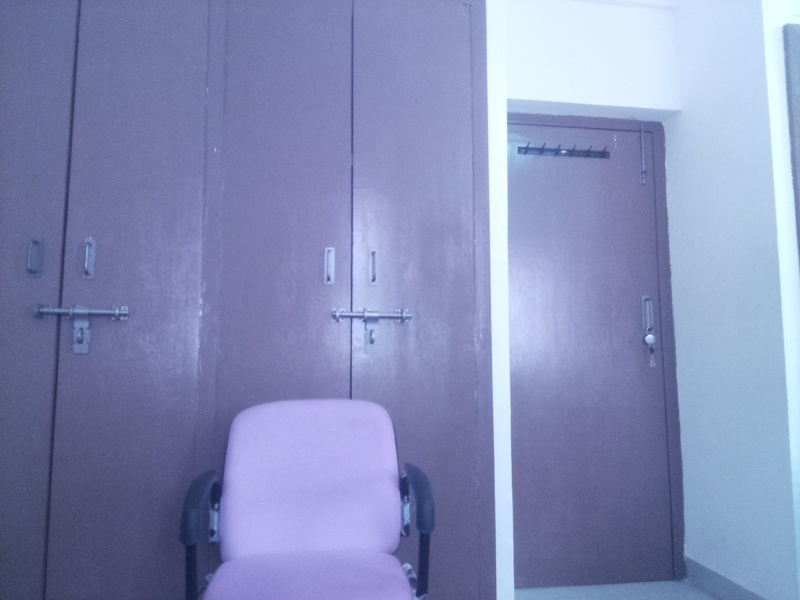} 
    \caption{} 
  \end{subfigure}
  \begin{subfigure}{.33\textwidth}
    \centering
    \includegraphics[width=.8\linewidth]{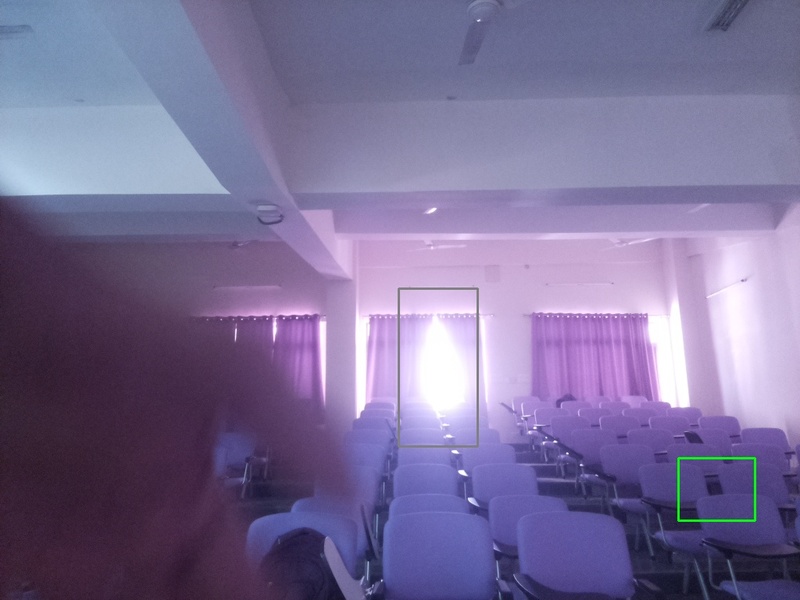}
    \caption{}
  \end{subfigure}
  \caption{Empty Room Images}
  \label{fig:empty}
\end{figure}

\begin{figure}
  \begin{subfigure}{.33\textwidth}
    \centering
    \includegraphics[width=.8\linewidth]{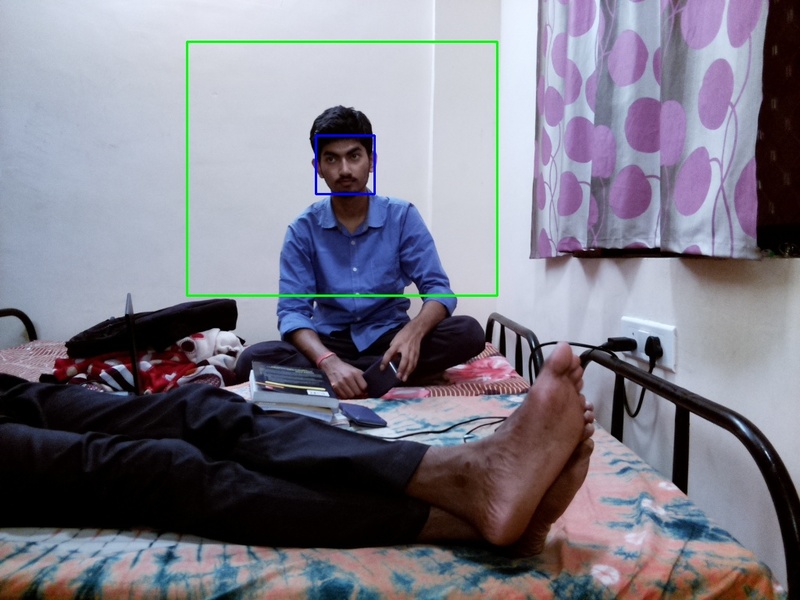} 
    \caption{} 
  \end{subfigure}
  \begin{subfigure}{.33\textwidth}
    \centering
    \includegraphics[width=.8\linewidth]{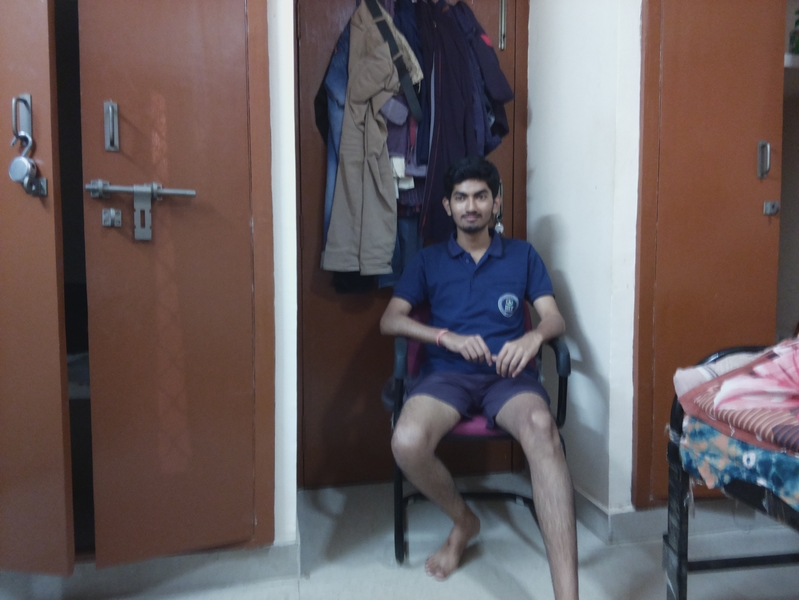} 
    \caption{} 
  \end{subfigure}
  \begin{subfigure}{.33\textwidth}
    \centering
    \includegraphics[width=.8\linewidth]{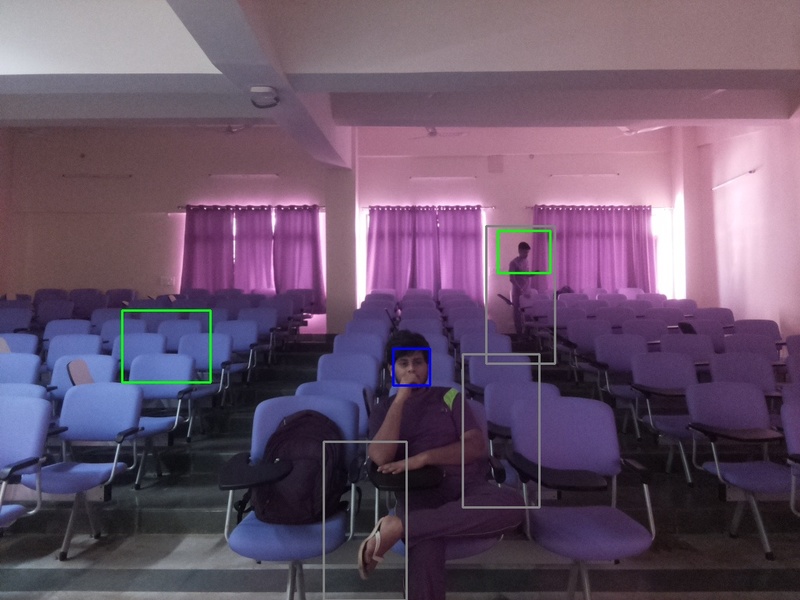}
    \caption{}
  \end{subfigure}
  \caption{Sitting Person Images}
  \label{fig:sitting}
\end{figure}

\begin{figure}
  \begin{subfigure}{.33\textwidth}
    \centering
    \includegraphics[width=.8\linewidth]{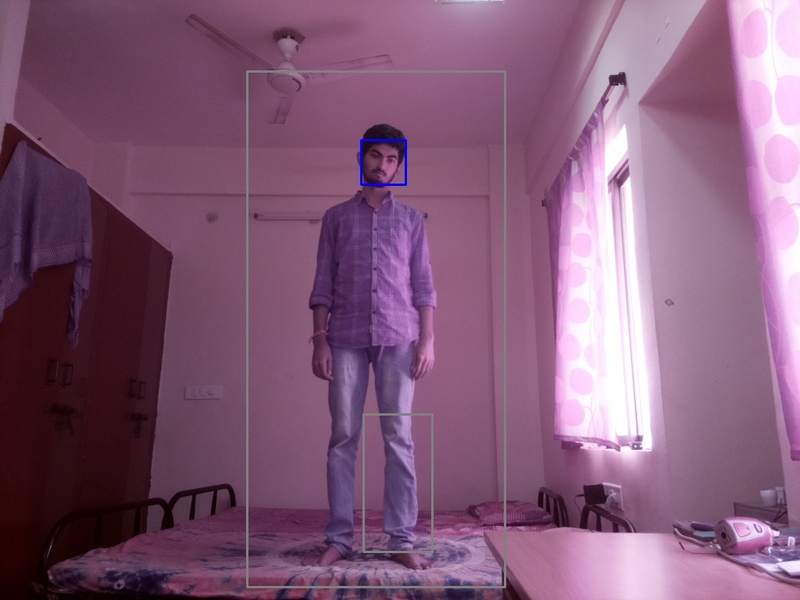} 
    \caption{} 
  \end{subfigure}
  \begin{subfigure}{.33\textwidth}
    \centering
    \includegraphics[width=.8\linewidth]{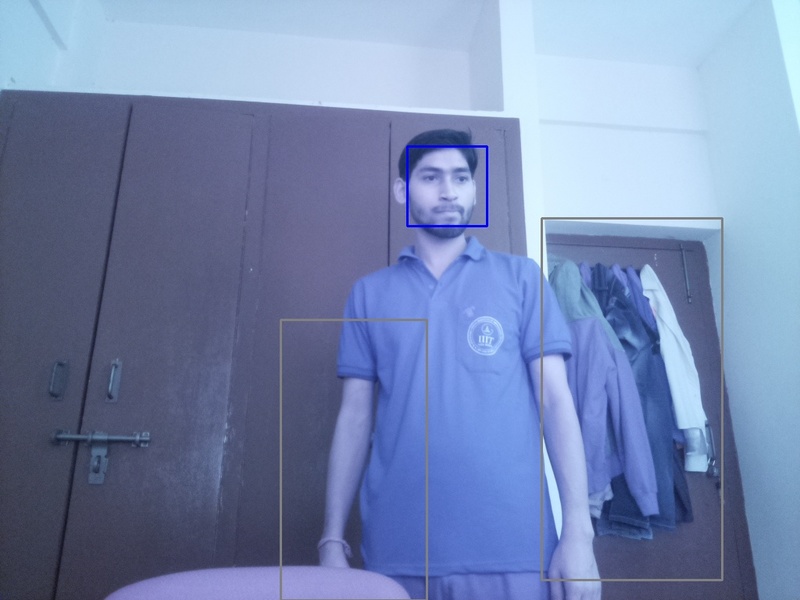} 
    \caption{} 
  \end{subfigure}
  \begin{subfigure}{.33\textwidth}
    \centering
    \includegraphics[width=.8\linewidth]{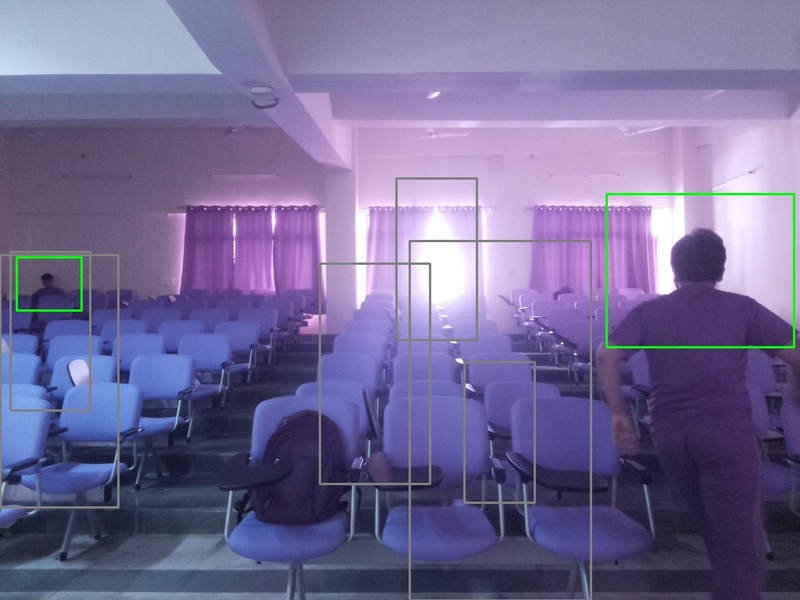}
    \caption{}
  \end{subfigure}
  \caption{Standing Person Images}
  \label{fig:standing}
\end{figure}

\begin{figure}
  \begin{subfigure}{.33\textwidth}
    \centering
    \includegraphics[width=.8\linewidth]{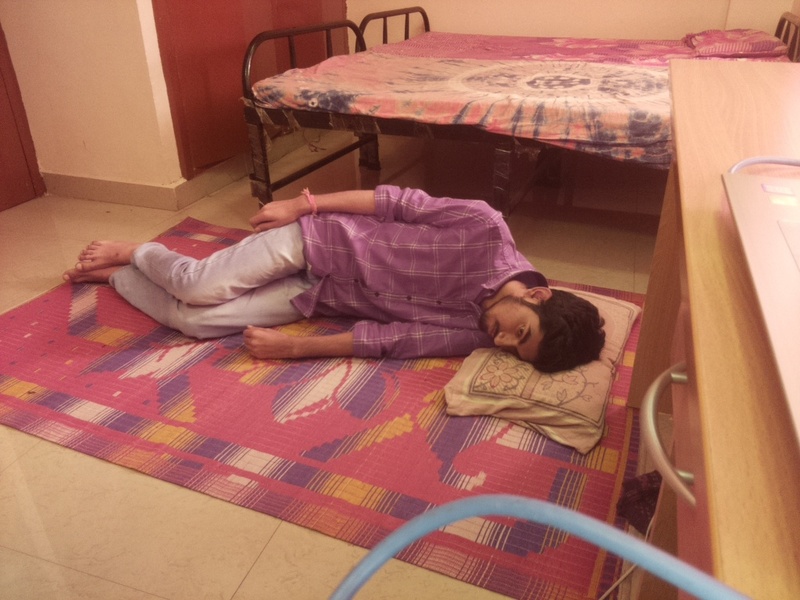} 
    \caption{} 
  \end{subfigure}
  \begin{subfigure}{.33\textwidth}
    \centering
    \includegraphics[width=.8\linewidth]{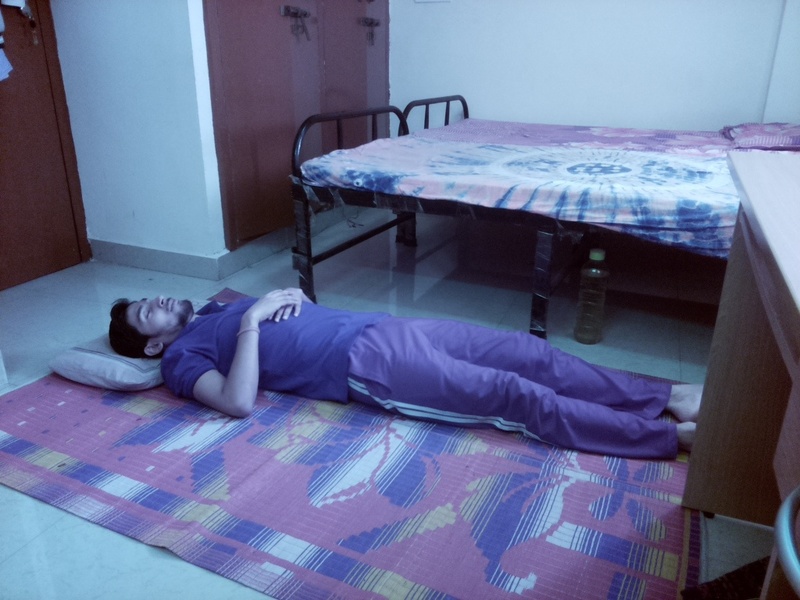} 
    \caption{} 
  \end{subfigure}
  \begin{subfigure}{.33\textwidth}
    \centering
    \includegraphics[width=.8\linewidth]{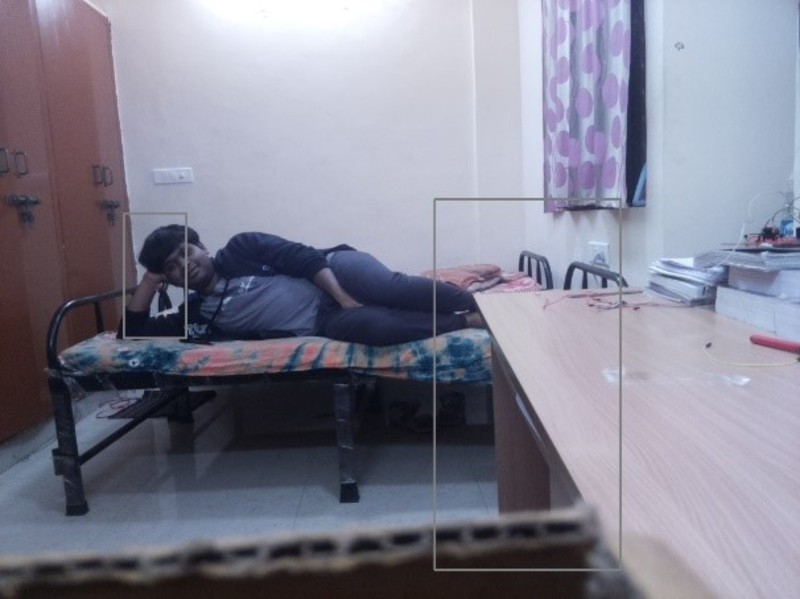}
    \caption{}
  \end{subfigure}
  \caption{Lying Person Images}
  \label{fig:lying}
\end{figure}

\begin{figure}
  \begin{subfigure}{.33\textwidth}
    \centering
    \includegraphics[width=.8\linewidth]{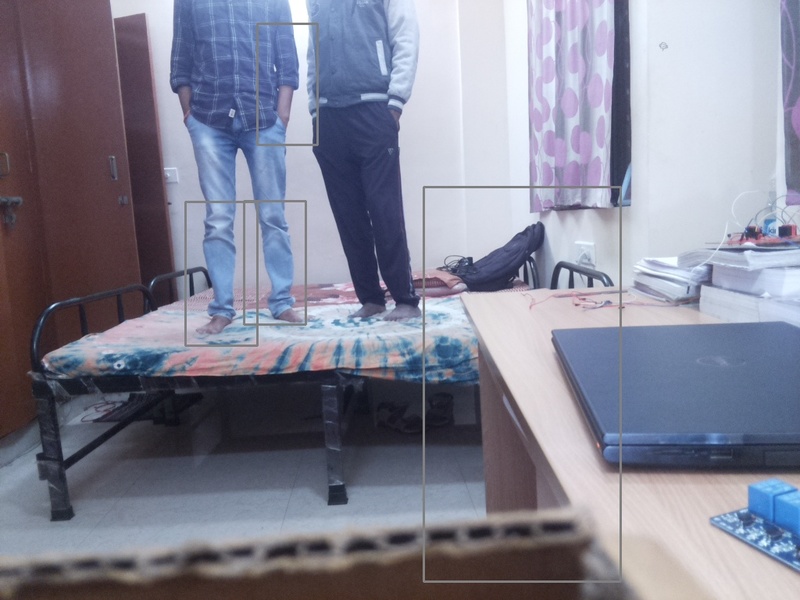} 
    \caption{} 
  \end{subfigure}
  \begin{subfigure}{.33\textwidth}
    \centering
    \includegraphics[width=.8\linewidth]{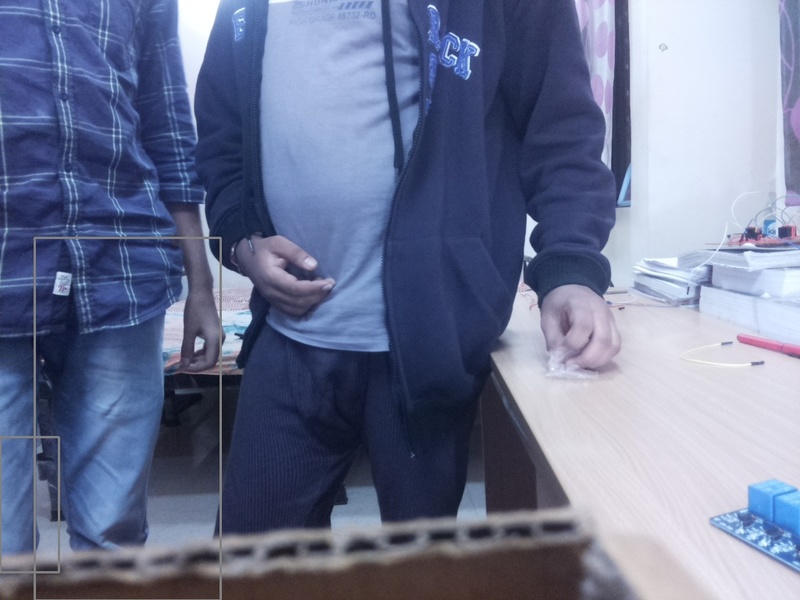} 
    \caption{} 
  \end{subfigure}
  \begin{subfigure}{.33\textwidth}
    \centering
    \includegraphics[width=.8\linewidth]{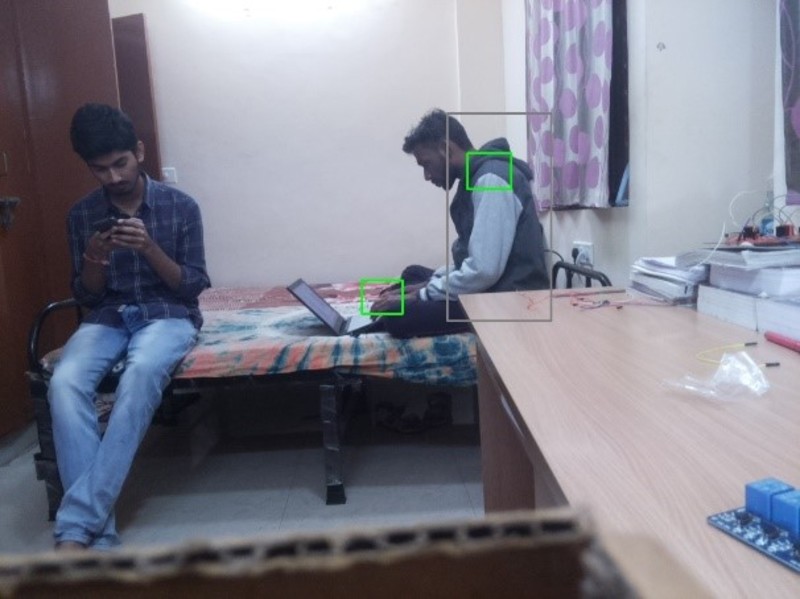}
    \caption{}
  \end{subfigure}
  \caption{Ambigious Images}
  \label{fig:ambigious}
\end{figure}

40 hours of controlled experiments were conducted in all. 10 hours of experiments were conducted in the lecture hall. Rest 30 hours were conducted in the dormitory in which 10 hours were conducted at night and 10 hours of experiments were conducted in day, 10 hours of experiments were conducted keeping the prototype at a different angle. The controlled experiments were determined according to the real-life situations arousing in a normal home. The experiments were categorised in the multiple sets based on states of the person in the room, empty room and non-clear images of the person and multiple people in the room. The first set of experiments was conducted of empty room where no person was visible in the field of vision of camera which simulated the real-life scenario of empty room. Figure \ref{fig:empty} shows the images of the empty room which was used captured and used to verify the results. This experiment was conducted by creating false positives responses from PIR sensor deliberately to get images of an empty room. This was done on purpose because of two reasons, first to simulate a false positive response from the PIR sensor which is a real-life problem. Secondly, to simulate a false motion in the absence of a human in the room which can be caused by a pet animal or an inanimate object. It also corresponds to the images of different rooms captured to remove the biases from learning the one room. Images were also captured at different angles and illumination. These varying illumination and angle are important for the prototype to classify in real-time deployment. The second set of experiments was conducted by people in states to be determined such as sitting, standing and lying. Figure \ref{fig:sitting} shows the images of the person sitting in the room. Figure \ref{fig:standing} shows the images of a person standing in the room. Figure \ref{fig:lying} shows the images of a person lying in the room. These experiments were explicitly conducted to determine the state of the person from the prototype. The third set of the experiments was conducted when the person or people were partially visible to the camera. Figure \ref{fig:ambigious} shows the non-clear images of a person standing in the room. These were simulated by the heuristic of camera angles. In a real life scenario, the deployment may have these set of images because the user roams in the room with his free will and does not move according to the camera position. The final set of experiments were conducted when there were multiple people in the room. We tried to conduct the experiments in different rooms with different people so as to remove any biases in the classification and validate our prototype in real-life scenario. Figure \ref{fig:ambigious} also shows the images of multiple people sitting in the room. These were simulated by making another person take the same state as the one who was before being monitored. This also justifies the real-life scenario which could include more than one person in the frame captured.

\section{Results and Analysis}
With the experiments conducted for various states, environments, people, illumination, angle and other parameters, we can safely say that our prototype is tested on all types of possible situation. The algorithm we propose HAUAR can distinctly classify between the four states (empty, sit, stand and lie) within 20 seconds. The proposed prototype also reduces the power consumption by not allowing the home appliances to be utilized in the absence of the user. It also optimally utilizes the home appliances in the presence of the user. The algorithm is also energy efficient as it does not allow classification to be done on the regular basis or capturing the image. This is achieved by making the vision module non-persistent and PIR Sensor to be persistent in detecting the motion. This reduces the energy consumption by the module. The solution also takes care of the issue of user data privacy by not streaming the images through any communication media or secure the images captured in any physical media. We have removed the intervention of human from our solution and had made the solution to be user-friendly by giving the user a choice to choose the intensity of light (dim or bright) and speed of the fan (low or high). This choice has been given due to a variety of user preferences in operating these appliances. Also, it was found in the experiment that HAAR cascade classifier and HOG classifier generates false-positive results due to which they cannot be depended upon for human detection. Table \ref{table:cm} summarises the results from experiments in the form of confusion matrix. Average accuracy of the model was calculated to be 90.35\%. Table \ref{table:compare} compares six different implemented home automation techniques with our proposed model. 

\begin{table}
  \caption{Confusion Matrix}
  \centering
  \begin{tabular}{lllll}
    \toprule
    States      & Empty Room  & Sitting     & Standing      & lying   \\
    \midrule
    Empty Room  & 586         & 0          & 0             & 14       \\
    Standing    & 0           & 957        & 143           & 0        \\
    Sitting     & 0           & 139        & 961           & 0        \\
    Lying       & 26          & 0          & 54            & 1020     \\
    \bottomrule
  \end{tabular}
  \label{table:cm}
\end{table}

\begin{table}
  \caption{Comparison of Our Proposed Model with 6 Different Implemented Home Automation Modules}
  \centering
  \begin{tabulary}{\linewidth}{JJJJJJJJ}
    \toprule
    Parameter & HAUAR & \cite{Starner} & \cite{Uhrikova} & \cite{Bentabet} & \cite{Deore} & \cite{Harsha} & \cite{Tsai} \\
    \midrule
    Technology & CV and IoT & CV & CV & EEG & IoT & IoT & CV and IoT \\
    Primary Detection & Action/ State of Person & Hand Gestures & Person Count & EEG Signals from Brain & No Detection & No detection & Human Height detection \\
    Primary Algorithm & HAUAR & Color Blob Detection and Hidden Markov Model & Background Subtraction MOG2 & P300 waves & N/A & N/A & Height Detection \\
    Running Platform & RPi & Remote PC & BBB & BCI & Arduino & RPi + Arduino & Transmission device with Bluetooth 4.0 \\
    Supporting Platform / Libraries & Tensorflow and OpenCV & N/A & Open CV & N/A & Web Interface & Android Smartphone & Android Smartphone \\
    Video/ Image Processing & Image & Image & Video & N/A & N/A & N/A & Video Surveillance \\
    Camera Position & Flexible & Only in Pendant & Only Overhead & N/A & N/A & N/A & Only in front of human \\
    Night Vision & Supported & Supported & Not Supported & N/A & N/A & N/A & Not Supported \\
    States which can be processed & Absent, Stand, Lie, and Sit & Four control and six user-defined gestures & Present or Absent & None & None & None & Adult or Child \\
    Human interaction & Not Required & Required & Not Required & Required & Required & Required & Required \\
    Primary Purpose & HA & Control Home Appliances & Security & Control Home Appliances & Control Home Appliances & Control Home Appliances & Control Home Appliances and Security \\
    Target Audience & All Public & General Public & General Public & Paralyzed People & General Public & General Public & General Public \\
    Training & Required & Required & Required & Required & Not Required & Not Required & Required \\
    Energy Consumption & Minimised & Low & High & N/A & N/A & High & N/A \\
    Intelligent & Yes & No & No & No & No & No & No \\
    Accuracy & 90\% & N/A & N/A & 88.66\% & N/A & N/A & 90\% \\
    Average Processing Time & 35 sec/frame & N/A & 0.12 sec/frame & Min 5 sec & N/A & N/A & N/A \\
    \bottomrule
	\end{tabulary}
  \label{table:compare}
\end{table}

\section{Conclusion}
With the result analysis and the comparative analysis of HAUAR with other implemented solutions. We can conclude that our solution (HAUAR) primarily focuses on the minimising human intervention as opposed to other solutions. We also have successfully deployed a nonhuman intervened system which is capable of taking decisions on its own based on classification and state table. Our solution also solves the problem of user data privacy by not storing the images captured. Also, as the system is self-sustained and does not require any connection to other devices and resources, we protect the system from breaching by isolating the device from the network. The information gathered and processing of information is also higher than the other implemented solutions but the comparative analysis shows a higher processing time than other implemented solutions. The prototype is also expected to help raise the user’s level of awareness of privacy and security in general for IoT environments, where securing sensitive user-generated information is an integral part. This project has been successfully tested on a single person and multiple people in the same state. Multiple people in the different state provide a different set of challenges such as occlusion between the persons, segmenting the image, bifurcating each people and incorporating a different state table which incorporates an equal number of people in state 1 and state 2 and also a number of people in state 1 then state 2. Incorporation of multiple people occupancy which is also an open research challenge in the domain of computer vision will be done as a future work on this project. 

\bibliographystyle{unsrt}

\begin{thebibliography}{1}

\bibitem{Walsh}
Walsh, R., \& Hornsby, A. (2011, January).\newblock Towards off-the-shelf computer vision for user interaction in consumer homes.\newblock In Consumer Electronics (ICCE), 2011 IEEE International Conference on (pp. 121-122). IEEE
\bibitem{Hossai}
Hossai, M. R. T., Shahjalal, M. A., \& Nuri, N. F. (2017, February).\newblock Design of an IoT based autonomous vehicle with the aid of computer vision.\newblock In Electrical, Computer and Communication Engineering (ECCE), International Conference on (pp. 752-756). IEEE. 
\bibitem{Geneiatakis}
Geneiatakis, D., Kounelis, I., Neisse, R., Nai-Fovino, I., Steri, G., \& Baldini, G. (2017, May).\newblock Security and privacy issues for an IoT based smart home.\newblock In Information and Communication Technology, Electronics and Microelectronics (MIPRO), 2017 40th International Convention on (pp. 1292-1297). IEEE. 
\bibitem{Jacobsson}
Jacobsson, A., \& Davidsson, P. (2015, December).\newblock Towards a model of privacy and security for smart homes.\newblock In Internet of Things (WF-IoT), 2015 IEEE 2nd World Forum on (pp. 727732). IEEE. 
\bibitem{Deore}
Deore, R. K., Sonawane, V. R., \& Satpute, P. H. (2015, December).\newblock Internet of Thing Based Home Appliances Control.\newblock In Computational Intelligence and Communication Networks (CICN), 2015 International Conference on (pp. 898-902). IEEE. 
\bibitem{Bentabet}
Bentabet, N., \& Berrached, N. E. (2016, November).\newblock Synchronous P300 based BCI to control home appliances. In Modelling, Identification and Control (ICMIC), 2016 8th International Conference on (pp. 835-838). IEEE. 
\bibitem{Kumar}
Kumar, P., \& Pati, U. C. (2016, November).\newblock Arduino and Raspberry Pi based smart communication and control of home appliance system.\newblock In Green Engineering and Technologies (ICGET), 2016 Online International Conference on (pp. 1-6). IEEE.
\bibitem{Patchava}
Patchava, V., Kandala, H. B., \& Babu, P. R. (2015, December).\newblock A Smart Home Automation technique with Raspberry Pi using IoT.\newblock In Smart Sensors and Systems (IC-SSS), International Conference on (pp. 1-4). IEEE. 
\bibitem{Peng}
Peng, Z., Kato, T., Takahashi, H., \& Kinoshita, T. (2015, October).\newblock Intelligent home security system using agent-based IoT Devices.\newblock In Consumer Electronics (GCCE), 2015 IEEE 4th Global Conference on (pp. 313-314). IEEE. 
\bibitem{Baroffio}
Baroffio, L., Bondi, L., Cesana, M., Redondi, A. E., \& Tagliasacchi, M. (2015, December).\newblock A visual sensor network for parking lot occupancy detection in Smart Cities.\newblock In Internet of Things (WF-IoT), 2015 IEEE 2nd World Forum on (pp. 745750). IEEE. 
\bibitem{Sefat}
Sefat, M. S., Khan, A. A. M., \& Shahjahan, M. (2014, May).\newblock Implementation of vision based intelligent home automation and security system.\newblock In Informatics, Electronics \& Vision (ICIEV), 2014 International Conference on (pp. 1-6). IEEE.  
\bibitem{Tsai}
Tsai, T. H., \& Zhang, K. L. (2016, October).\newblock Implementation of intelligent home appliances based on IoT.\newblock In Circuits and Systems (APCCAS), 2016 IEEE Asia Pacific Conference on (pp. 146148). IEEE.  
\bibitem{Shinde}
Shinde, R. V., Shimpi, S. M., Lanjewar, P. S., Nivangune, P. A., Mundkar, D. S., \& Sonawane, A. R. (2016).\newblock Vision Based Hand Gesture Recognition for Real Time Home Automation Application.\newblock International Journal of Engineering Science, 3176.  
\bibitem{Starner}
Starner, T., Auxier, J., Ashbrook, D., \& Gandy, M. (2000, October).\newblock The gesture pendant: A self-illuminating, wearable, infrared computer vision system for home automation control and medical monitoring.\newblock In Wearable computers, the fourth international symposium on (pp. 87-94). IEEE.  
\bibitem{Harsha}
Harsha, S. S., Reddy, S. C., \& Mary, S. P. (2017, February).\newblock Enhanced home automation system using the Internet of Things.\newblock In I-SMAC (IoT in Social, Mobile, Analytics, and Cloud)(ISMAC), 2017 International Conference on (pp. 89-93). IEEE. 
\bibitem{Uhrikova}
Uhrikova, Z., Nugent, C. D., \& Hlavac, V. (2008, August). The use of computer vision techniques to augment home based sensorised environments. In Engineering in Medicine and Biology Society, 2008. EMBS 2008. 30th Annual International Conference of the IEEE (pp. 2550- 2553). IEEE.
\end{thebibliography}

\end{document}